\documentclass[aps,prd,reprint,longbibliography,titlepage,nofootinbib]{revtex4-2}

\usepackage{bm}
\usepackage[utf8]{inputenc}
\usepackage[T1]{fontenc}
\usepackage{amsmath}
\usepackage{amsfonts}
\usepackage{mathtools} 
\usepackage{enumerate}
\usepackage{hyperref}
\hypersetup{colorlinks=true,linkcolor=blue,urlcolor=blue,citecolor=blue}
\usepackage[cal=boondox]{mathalfa}
\usepackage{lipsum}
\usepackage{relsize}
\usepackage{color}
\usepackage{nameref}

\usepackage{soul}

\begin{document}
	
\title{Diffeomorphism-invariant action principles for trace-free Einstein gravity}
		
\author{Merced Montesinos\,\href{https://orcid.org/0000-0002-4936-9170} {\includegraphics[scale=0.05]{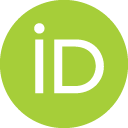}}}
\email{merced@fis.cinvestav.mx}

\affiliation{Departamento de F\'{i}sica, Cinvestav, Avenida Instituto Polit\'{e}cnico Nacional 2508, San Pedro Zacatenco,\\
07360 Gustavo A. Madero, Ciudad de M\'exico, M\'exico}

\author{Diego Gonzalez\,\href{https://orcid.org/0000-0002-0206-7378} {\includegraphics[scale=0.05]{ORCIDiD_icon128x128.png}}}
\email{dgonzalezv@ipn.mx}
 
\affiliation{Departamento de F\'{i}sica, Cinvestav, Avenida Instituto Polit\'{e}cnico Nacional 2508, San Pedro Zacatenco,\\
07360 Gustavo A. Madero, Ciudad de M\'exico, M\'exico}
\affiliation{Escuela Superior de Ingenier\'ia Mec\'anica y El\'ectrica, Instituto Polit\'ecnico Nacional, 07738 Ciudad de M\'exico, M\'exico}
 
\date{\today}
	
\begin{abstract}
Trace-free Einstein gravity is a theory of gravity that is an alternative to general relativity, wherein the cosmological constant arises as an integration constant. However, there are no fully diffeomorphism-invariant action principles available that lead to the equations of motion of this theory. Unimodular gravity comes close to this idea, but it relies on action principles that are invariant only under volume-preserving diffeomorphisms. We present a real $BF$-type action principle for trace-free Einstein gravity that is fully diffeomorphism-invariant and does not require any unimodular condition or nondynamical fields. We generalize this action principle by giving another one involving a free parameter. 
\end{abstract}

\maketitle

\section{Introduction}
After formulating general relativity, Einstein himself proposed an alternative theory of gravity~\cite{Einstein_1919,Einstein_1952} in which the equations of general relativity are replaced by their trace-free part~\cite{Einstein_1927}
\begin{eqnarray}\label{TFEM}
R_{\mu\nu} - \frac14 R g_{\mu\nu} =\frac{8 \pi G}{c^4} \Big( T_{\mu\nu} - \frac14 T g_{\mu\nu}\Big).
\end{eqnarray} 
Here, the Ricci tensor $R_{\mu\nu}$ and the scalar curvature $R$ are constructed out of the Levi-Civita connection compatible with the metric $g_{\mu\nu}$ and $T_{\mu\nu}$ is the energy-momentum tensor. These equations, supplemented with the assumption of the conservation of the energy-momentum, yield the same consistent results of Einstein's general relativity~\cite{Ellis_2011,Ellis_2014}; however, with the advantage that the cosmological constant now naturally emerges as an integration constant implied by the second Bianchi identity, thus solving the fundamental issue of vacuum energy~\cite{Ellis_2019}. Despite this remarkable property, there is no completely diffeomorphism-invariant action principle that leads to the trace-free Einstein equations~\eqref{TFEM}.  An approach that almost realizes this idea is unimodular gravity~\cite{Anderson_1971,Weinberg_1989}. In this theory, equations~\eqref{TFEM} are obtained from an action principle that is not totally diffeomorphism-invariant, but only invariant under volume-preserving diffeomorphisms due to a unimodular condition that fixes the determinant of the metric. Another related action principle is that of Ref.~\cite{Henneaux1989}. However, while such an action is diffeomorphism-invariant, it does not yield the trace-free Einstein equations; instead, it gives the equations of general relativity with a generalized version of the unimodular condition~\cite{JackvanDam} and a cosmological constant that arises as an integration constant via a Lagrange multiplier~\cite{Henneaux1989,Carballo-Rubio_2022}. Furthermore, in this case, the origin of the cosmological constant is unrelated to the second Bianchi identity. 

The goal of this paper is to address this longstanding problem by providing the first fully diffeomorphism-invariant action principle for trace-free Einstein gravity. Our action principle expresses trace-free Einstein gravity as a real constrained $BF$ theory, without involving the unimodular condition or requiring any nondynamical fields, thus preserving the full diffeomorphism invariance while retaining the spirit of Einstein's proposal~\cite{Einstein_1919,Einstein_1952,Einstein_1927} in the sense that the cosmological constant arises from the second Bianchi identity. We also present a generalization of this $BF$-type action that introduces a free parameter without modifying the equations of motion. A characteristic feature of our action principles is that they not only describe trace-free Einstein gravity, but also contain a gravitational sector corresponding to Einstein's general relativity. Thus, our work opens a new research avenue to study the quantum aspects of trace-free Einstein gravity, in particular in the approach to quantum gravity known as spinfoam models~\cite{RovBook,ThieBook,Perez_2013}, where $BF$-type theories~\cite{pleb1977118,BFgravity} play a fundamental role. This can provide an opportunity for comparison with previous results on quantum unimodular gravity~\cite{Smolin_2009,Smolin_2011}.

We will first introduce a first-order formulation of~\eqref{TFEM} in terms of 2-forms and then prove that the equations of motion arising from our action principles are equivalent to the trace-free Einstein equations in the first-order formalism. Finally, we present a discussion  of our results, drawing out some implications and outlining future work.

\section{Trace-free Einstein gravity in the tetrad and connection formalism}
In the first-order formalism of gravity, the fundamental variables are an orthonormal frame of 1-forms $e^I$, where $I=0,1,2,3$, and a spacetime connection $\omega^I{}_J$ compatible with the metric $(\eta_{IJ}) = \mbox{diag} (\sigma,1,1,1)$ ($\sigma=-1$ for the Lorenzian case and $\sigma=1$ for the Euclidean one) and torsion-free, i.e., $D \eta_{IJ}:= d \eta_{IJ} - \omega^K{}_I \eta_{KJ} - \omega^K{}_J \eta_{IK}=0$ and
\begin{equation}\label{spinc}
D e^I := d e^I + \omega^I{}_J \wedge e^J=0,
\end{equation}
where $d$ is the exterior derivative and $\wedge$ is the wedge product. The connection $\omega^I{}_J$ is referred to as the torsion-free spin connection, and its curvature is given by $\mathcal{R}^I{}_J = d \omega^I{}_J + \omega^I{}_K \wedge \omega^K{}_J$. Furthermore, we take $T_{IJ}$ as the components of the energy-momentum tensor in the orthonormal frame, from which we introduce the object $T_I = T_{IJ} e^J$. Additionally, $\varepsilon_{IJKL}$ is the totally antisymmetric $SO(3,1)$ invariant tensor if $\sigma=-1$ [respectively, $SO(4)$ if $\sigma=1$] with $\varepsilon_{0123}=1$. Capital indices from the middle of the alphabet $I, J, K, \ldots$ are raised (lowered) with $\eta^{IJ}$ ($\eta_{IJ}$). 

The trace-free Einstein equations with matter fields~\eqref{TFEM} in the orthonormal frame are
\begin{eqnarray}\label{eq_forms}
\ast \mathcal{R}_{IJ}-\star \mathcal{R}_{IJ}=\frac{8 \pi G}{c^4} \Big ( \ast \mathcal{T}_{IJ}-\star \mathcal{T}_{IJ} \Big),
\end{eqnarray}
where $\mathcal{T}_{IJ}:=\frac12 (T_I \wedge e_J - T_J \wedge e_I)$, $\ast$ stands for the internal dual and so $\ast \mathcal{R}_{IJ}= \frac12 \varepsilon_{IJ}{}^{KL} \mathcal{R}_{KL}$ and 
$\ast \mathcal{T}_{IJ} =\frac12 \varepsilon_{IJ}{}^{KL} \mathcal{T}_{KL}$, 
and $\star$ is the Hodge dual. 

In fact, expanding the curvature in the orthonormal frame, $\mathcal R_{IJ} =\frac12 R_{IJKL} e^K \wedge e^L$, and using $\star (e^I \wedge e^J)= \frac12 \varepsilon^{IJ}{}_{KL} e^K \wedge e^L$ to compute $\star \mathcal{R}_{IJ}$ and $\star \mathcal{T}_{IJ}$, it is easy to show that~\eqref{eq_forms} is equivalent to 
\begin{eqnarray}\label{TFEM2}
R_{IJ} -\frac 14 R \eta_{IJ}=\frac{8 \pi G}{c^4} \Big ( T_{IJ} - \frac14 T \eta_{IJ} \Big ),
\end{eqnarray}
which are the equations~\eqref{TFEM} in the orthonormal frame. Tensor equations~\eqref{eq_forms} hold in any frame, of course. 

The trace-free Einstein equations~\eqref{eq_forms} carry a deep geometric and algebraic meaning: they reveal a remarkable equality between the internal and Hodge duals of $\mathcal{R}_{IJ}-\frac{8 \pi G}{c^4}\mathcal{T}_{IJ}$. In this sense, it is remarkable that the formulation~\eqref{eq_forms}--expressed in terms of 2-forms--can only be defined in four-dimensional spacetimes. This is in contrast to the definition of the theory in the metric formalism~\eqref{TFEM}, which can be extended to $n$-dimensional spacetimes.

Equations~\eqref{spinc} and~\eqref{eq_forms} must be supplemented with the equations for the matter fields. Additionally, we need to specify whether the energy-momentum is conserved or not (see Refs.~\cite{Ellis_2011,Ellis_2014,PhysRevLett.118.021102} for some analyses in the metric formalism). Here, we assume that the energy-momentum is conserved. Therefore, using the second Bianchi identity, the conservation of the energy-momentum, and~\eqref{TFEM2}, the cosmological constant $\Lambda$ emerges as an integration constant
\begin{eqnarray}
R + \frac{8 \pi G}{c^4} T = 4 \Lambda. 
\end{eqnarray}
Additionally, we can construct out another version of trace-free Einstein gravity in the first-order formalism, in which the conservation of the energy-momentum is not assumed.  This alternative version, which allows for the violation of the energy-momentum, differs from the one of Ref.~\cite{PhysRevLett.118.021102} in the metric formalism because it does not involve the unimodular condition.

It is also important to remark that trace-free Einstein gravity in the first-order formalism, given by~\eqref{spinc} and~\eqref{eq_forms}, is totally diffeomorphism-covariant and also covariant under $SO(3,1)$ [or $SO(4)$] transformations of the frame and the connection. This is because no condition on the volume (i.e., on the tetrad field $e^I$) is imposed, as is the case in unimodular gravity in the first-order formalism~\cite{PhysRevD.92.024036}, meaning that in the formulation~\eqref{spinc} and~\eqref{eq_forms} the orthonormal frame $e^I$ involves sixteen independent components. Incidentally, an action principle totally diffeomorphism-invariant involving the sixteen independent components of the tetrad $e^I$, the connection $\omega^I{}_J$, and the matter fields that leads to~\eqref{spinc} and~\eqref{eq_forms}, and to the equations of the matter fields, is still missing.  

\section{Action principle}
We now turn to the $BF$-type action for trace-free Einstein gravity, which is the central result of this paper. The fully diffeomorphism-invariant action principle $S[A,B,\phi,\mu]$ is given by
\begin{align}\label{BF_TFG}
S=& \int \bigg[B^{IJ} \wedge {\mathcal F}_{IJ}- \frac12 \left ( {\ast\phi}_{IJKL} + {\phi\ast}_{IJKL} \right ) B^{IJ} \wedge B^{KL} \nonumber \\
&- \mu \phi^{IJ}{}_{IJ}\bigg],
\end{align}
where $B^{IJ}$ is a 2-form  fulfilling $B^{IJ}=-B^{JI}$, ${\mathcal F}^I{}_J = d A^I{}_J + A^I{}_K \wedge A^K{}_J$ is the curvature of the connection $A^I{}_J$ satisfying $A^{IJ}=-A^{JI}$, the field $\phi_{IJKL}$ has the symmetries $\phi_{IJKL}=-\phi_{JIKL}=-\phi_{IJLK}=\phi_{KLIJ}$, and $\mu$ is a 4-form.  Furthermore, ${\ast \phi}_{IJKL} = \frac12 \varepsilon_{IJ}{}^{MN} \phi_{MNKL}$ and ${\phi \ast}_{IJKL} = \frac12 \phi_{IJMN} \varepsilon^{MN}{}_{KL}$ are the left and right internal duals of $\phi_{IJKL}$, respectively.

We now show that the trace-free Einstein equations come from this action principle. The variation of~\eqref{BF_TFG} with respect to $\phi_{IJKL}$, $A_{IJ}$, $B^{IJ}$, and $\mu$ yields, respectively, the equations of motion:
 \begin{subequations}
 \begin{align}
 & {\ast B}^{IJ} \wedge B^{KL} + B^{IJ} \wedge {\ast B}^{KL} \nonumber\\
 & + \mu (\eta^{IK}\eta^{JL}-\eta^{IL}\eta^{JK})=0, \label{varauxfield}\\
 & D B^{IJ}=0, \label{varA} \\
 & {\mathcal F}_{IJ} = \left ( {\ast\phi}_{IJKL} + {\phi\ast}_{IJKL} \right ) B^{KL}, \label{varB}\\
  & \phi^{IJ}{}_{IJ}=0. \label{varmu}
 \end{align}     
 \end{subequations}

After getting rid of $\mu$,~\eqref{varauxfield} has the only solutions
\begin{align}
B^{IJ} &= k_1 e^I \wedge e^J, \label{first_sector} \\
B^{IJ} &= k_2 \ast \left ( e^I \wedge e^J \right ), \label{second_sector}
\end{align}
where $e^I$ is a set of four linearly independent 1-forms and $k_1$ and $k_2$ are real nonvanishing constants. Then, \eqref{varA} and either~\eqref{first_sector} or~\eqref{second_sector} imply that $A^I{}_J$ is the torsion-free spin connection. 

We now prove that sector~\eqref{first_sector} is trace-free Einstein gravity.  Expanding the curvature as ${\mathcal F}^I{}_J = \frac12 F^I{}_{JKL} e^K \wedge e^L$ and using~\eqref{first_sector},~\eqref{varB} amounts to $F_{IJKL} = 2 k_1 \left ( {\ast\phi}_{IJKL} + {\phi\ast}_{IJKL} \right )$. This relation and the first Bianchi identity, ${\mathcal F}^I{}_J \wedge e^J=0$, imply
\begin{equation}\label{TFEM1}
 F_{IJ}=k_1 {\ast\phi}^{KL}{}_{KL} \eta_{IJ}, \end{equation}
with $F_{IJ} = F^K{}_{IKJ}$. Getting rid of ${\ast\phi}^{KL}{}_{KL}$ in this relation, it becomes 
\begin{eqnarray}\label{TFEM2*}
F_{IJ} -\frac 14 F \eta_{IJ}=0,
\end{eqnarray}
with $F= F^{IJ}{}_{IJ}$. Equations~\eqref{TFEM2*} are precisely the trace-free Einstein equations \eqref{TFEM2} in the absence of matter fields.

Along the same lines, we can show that sector~\eqref{second_sector} corresponds to general relativity without cosmological constant. In fact, from~\eqref{varB} and~\eqref{second_sector} we have $F_{IJKL}= 2 k_2 \left ( {\ast\phi\ast}_{IJKL} + \sigma{\phi}_{IJKL} \right )$, which together with the first Bianchi identity imply $F_{IJ}=\sigma k_2 \phi^{KL}{}_{KL} \eta_{IJ}$. Then, using~\eqref{varmu}, it follows that 
\begin{eqnarray}\label{EGR}
F_{IJ}=0.
\end{eqnarray}
These are Einstein's equations for general relativity without cosmological constant. 

Thus, we have shown that the action principle~\eqref{BF_TFG} involves two distinct gravitational sectors, one of which is trace-free Einstein gravity. In fact, our analysis shows that sector~\eqref{first_sector} gives rise to trace-free Einstein equations~\eqref{TFEM2*}. Therefore, the formulation~\eqref{BF_TFG} is in total agreement with the (diffeomorphism-covariant) trace-free Einstein equations in the sense that the action principle~\eqref{BF_TFG} is totally diffeomorphism-invariant. On the other hand, Einstein's general relativity without cosmological constant emerges from sector~\eqref{second_sector}. Action principles having different sectors are not new. Among them one has, for instance, the $BF$-type action of Ref.~\cite{Pietri_1999}, which includes Einstein's general relativity and a topological sector.

Note that by performing the transformation $\phi_{IJKL} \rightarrow \sigma{\phi\ast}_{IJKL}$ and $\mu \rightarrow 2\sigma\mu$ in the action~\eqref{BF_TFG}, it acquires the equivalent form 
\begin{align}\label{BF_TFG_alt}
S=& \int \Big[ B^{IJ} \wedge {\mathcal F}_{IJ}- \frac12 \left ( \phi_{IJKL} + \sigma {\ast\phi\ast}_{IJKL} \right ) B^{IJ} \wedge B^{KL} \nonumber \\
&- \mu \epsilon^{IJKL} \phi_{IJKL}\Big],
\end{align}
where ${\ast \phi \ast}_{IJKL} = \frac14 \varepsilon_{IJ}{}^{MN} \phi_{MNPQ} \varepsilon^{PQ}{}_{KL}$.

\section{Generalized action involving a parameter}
A totally diffeomorphism-invariant action principle $S[A,B,\phi,\mu]$ that generalizes~\eqref{BF_TFG} is given by
\begin{align} \label{ActionPar}
S=& \int \Big[\left(B^{IJ}+\gamma {\ast B}^{IJ}\right) \wedge {\mathcal F}_{IJ} \nonumber \\
&- \frac12 \left ( {\ast\phi}_{IJKL} + {\phi\ast}_{IJKL} \right ) B^{IJ} \wedge B^{KL} \nonumber \\
&- \mu \Big (\left(1+\sigma \gamma^2\right) \phi^{IJ}{}_{IJ}-\gamma\epsilon^{IJKL} \phi_{IJKL} \Big )\Big],
\end{align}
where $\gamma$ is a real parameter. Note that in the case  $\gamma=0$ this action simply reduces to~\eqref{BF_TFG}. Remarkably, trace-free Einstein gravity also emerges from~\eqref{ActionPar} when $\gamma\neq0$, which is another important result of this paper.

We now show that~\eqref{ActionPar} is indeed an action principle for trace-free Einstein gravity when $\gamma\neq0$. The equation of motion coming from the variation of~\eqref{ActionPar} with respect to $\phi_{IJKL}$ implies that $B^{IJ}$ has the solutions
\begin{align}
B^{IJ} &= k_1 \Big (e^I \wedge e^J -\gamma \ast \left ( e^I \wedge e^J \right )  \Big ), \label{first_sector_Par} \\
B^{IJ} &= k_2 \Big  (e^I \wedge e^J -\frac{\sigma}{\gamma} \ast \left ( e^I \wedge e^J \right ) \Big ), \label{second_sector_Par}
\end{align}
with $k_1$ and $k_2$ real parameters. Then, the equation of motion for $A_{IJ}$ and either \eqref{first_sector_Par} or \eqref{second_sector_Par} state that $A^I{}_J$  is the torsion-free spin connection, provided that $\gamma^2\neq\sigma$. 

 We now focus on sector~\eqref{first_sector_Par}. The use of~\eqref{first_sector_Par}, the equation of motion obtained from the variation of~\eqref{ActionPar} with respect to $B^{IJ}$, and the first Bianchi identity imply
\begin{equation}\label{TFE-gral}
 F_{IJ}= k_1 \frac{\left(1-\sigma \gamma^2\right)}{\left(1+\sigma \gamma^2\right)}  {\ast\phi}^{KL}{}_{KL} \eta_{IJ}.  
\end{equation}
After getting rid of ${\ast\phi}^{KL}{}_{KL}$ in~\eqref{TFE-gral}, it becomes the trace-free Einstein equations~\eqref{TFEM2*}. Therefore, we conclude that sector~\eqref{first_sector_Par} is trace-free Einstein gravity. Following an analogous procedure, it can be shown that sector \eqref{second_sector_Par} leads to Einstein's equations for general relativity~\eqref{EGR}. 

Thus, it is noteworthy that the presence of the terms containing the parameter~$\gamma$ in~\eqref{ActionPar} does not modify the equations of motion~\eqref{TFEM2*} or~\eqref{EGR}, but allows the introduction of $\gamma$ as a free parameter into the action~\eqref{ActionPar}. In this sense, the action~\eqref{ActionPar} resembles the $BF$-type action for general relativity of Ref.~\cite{CMPR} (see also~\cite{BFgravity}), which is equivalent to the Holst action~\cite{Holst9605}, where the addition of the so-called Holst term to the Palatini action does not change the equations of motion, but introduces a free parameter into the Lagrangian for general relativity.

\section{Conclusions}
In this paper, we have presented the first real $BF$-type action principles for trace-free Einstein gravity, which are fully diffeomorphism-invariant. The main features of our action principles can be summarized as follows: (i)~All the fundamental variables involved in the actions are dynamical. (ii)~The trace-free Einstein equations arise from the actions without the need to impose any unimodular condition. Note that (i) and (ii) are themselves nontrivial properties, which are in striking contrast to the current paradigm that states that a formulation of the trace-free Einstein equations must necessarily involve a unimodular condition or a nondynamical field. (iii)~The cosmological constant emerges as an integration constant from the geometric structure of the theory through the use of the second Bianchi identity, in perfect agreement with the theoretical framework of Einstein's proposal~\cite{Einstein_1919,Einstein_1952,Einstein_1927}. (iv)~In addition to trace-free Einstein gravity, the actions involve a gravitational sector corresponding to Einstein's general relativity. There are no topological sectors in our action principles. (v)~The generalized action~\eqref{ActionPar} allows the inclusion of a free parameter, without modifying the equations of motion of trace-free Einstein gravity (and Einstein's general relativity). 

Apart from the action principles previously presented, we report that, in the Lorentzian case ($\sigma=-1$), the totally diffeomorphism-invariant action principle $S[A,B,\phi,\mu]$ given by
\begin{align} \label{ActionAlt}
S=& \int \Big[\left(B^{IJ} - {\ast B}^{IJ}\right) \wedge {\mathcal F}_{IJ} \nonumber \\
&- \frac12 \left ( {\ast\phi}_{IJKL} + {\phi\ast}_{IJKL} \right ) B^{IJ} \wedge B^{KL} \nonumber \\
&- \mu \epsilon^{IJKL} \phi_{IJKL} \Big],
\end{align}
also has two gravitational sectors: one leading to the trace-free Einstein equations~\eqref{TFEM2*} and one leading to Einstein's equations for general relativity~\eqref{EGR}. Clearly, this action principle also possesses the aforementioned properties.

The action principles obtained in this paper motivate applications in a wide range of areas in physics. In the context of loop quantum gravity, since our action principles correspond to a $BF$ theory with constraints on the $B$ field, they serve as a very suitable starting point for the (path integral) quantization of trace-free Einstein within the program of spinfoam models~\cite{Perez_2013}. In this context, the Hamiltonian formulation of these action principles is crucial for their canonical quantization, in particular to determine whether the techniques of loop quantum gravity~\cite{RovBook,ThieBook} can be applied, where diffeomorphism invariance is fundamental. Furthermore, it would be important to investigate whether the free parameter $\gamma$, included in the action~\eqref{ActionPar}, plays a relevant role at the quantum level. In the context of cosmology, trace-free Einstein gravity gives the same practical results as Einstein's general relativity, with the advantage that $\Lambda$ is now an integration constant~\cite{Ellis_2011,Ellis_2014}.  

Another noteworthy aspect of our action principles is that they allow for a variety of generalizations. First, notice that in our analysis we have only considered nondegenerate sectors (${\ast B}_{IJ} \wedge B^{IJ}\neq0$). It would also be relevant to analyze degenerate sectors, particularly in the quantum realm where nontrivial effects could arise. Additionally, since trace-free Einstein gravity in the metric formalism is not limited to any specific spacetime dimension, it would also be very valuable to extend our $BF$-type formulations to dimensions other than four. This would open up the possibility of applying the spinfoam quantization procedure to higher-dimensional trace-free Einstein gravity. Among other possible generalizations, a natural one is to extend our action principles to include the couplings of matter fields to trace-free Einstein gravity. In particular, it would be interesting to investigate the role of the free parameter $\gamma$  involved in the action~\eqref{ActionPar} when matter fields are included. Lastly, notice that if nonvanishing torsion is allowed within the theoretical framework, the resulting theory will differ from trace-free Einstein gravity, in the same sense that the Einstein-Cartan theory differs from Einstein's general relativity. In this regard, we also note that the fully diffeomorphism-invariant action principle $S[e,\omega]=\int R^2 \eta$, where $\eta=\frac{1}{4!}\epsilon_{IJKL} e^I \wedge e^J \wedge e^K \wedge e^L$ is the volume form, yields both (gravitationally induced) nonvanishing torsion and the trace-free Einstein equations in the first-order formalism without involving any unimodular condition~\cite{MontesinosfR_2020}.

Finally, the action principles presented in this paper can be modified to obtain only the equations of motion of trace-free Einstein gravity, eliminating the general relativity 
sector. This can be achieved by requiring that the 4-form $\mu$ in the action principles is nondynamical (which amounts to fix the volume). In this case, only the equations of motion of trace-free Einstein gravity arise but the tetrads satisfy the unimodular condition. A detailed analysis of this will appear elsewhere. 

\acknowledgments
Warm thanks to Alejandro Perez, Carlo Rovelli, and Ulises Nucamendi for their valuable comments. Diego Gonzalez acknowledges the financial support of Instituto Polit\'ecnico Nacional, Grant No. SIP-20230323.

\bibliography{references}
	
\end{document}